\documentclass[conference]{IEEEtran}
\ifCLASSINFOpdf
\else
\fi
\usepackage[cmex10]{amsmath}
\usepackage{graphicx}
\begin{document}
\title{Flexible  Authentication in Vehicular Ad hoc Networks}
\author{\IEEEauthorblockN{P. Caballero-Gil, C. Caballero-Gil, J. Molina-Gil and C. Hern\'andez-Goya}
\IEEEauthorblockA{Department of Statistics, Operations Research and
Computing University of La Laguna\\
38271 La Laguna. Tenerife.Spain.\\
Email:\{pcaballe, candidoc, jmmolina, mchgoya\}@ull.es}}
\maketitle
\begin{abstract}

A Vehicular Ad-Hoc Network (VANET) is a form of Mobile ad-hoc network,
to provide communications among nearby vehicles and between vehicles and
nearby fixed roadside equipment.
The key operation in VANETs is the broadcast of messages. Consequently, the vehicles need to make
sure that the information has been sent by an authentic node in the network.
VANETs present unique challenges such as high node mobility, real-time constraints, scalability, gradual deployment and privacy. No existent technique addresses all these requirements. In particular, both inter-vehicle and vehicle-to-roadside wireless communications present different characteristics that should be taken into account when defining node authentication services. That is exactly what is done in this paper, where the features of inter-vehicle and vehicle-to-roadside communications are analyzed to propose differentiated services for node authentication, according to privacy and efficiency needs.

\end{abstract}
\section{Introduction}
\footnotetext{Work developed in the frame of the project TIN2008-02236/TSI supported by the Spanish Ministry of Science and Innovation and FEDER Funds.\\
Proceedings of APCC IEEE Asia Pacific Conference on Communications. Vol. 208 , (October 2009)	pp. 876-879.}

\begin{figure}
 \centering
 \includegraphics[scale=0.55]{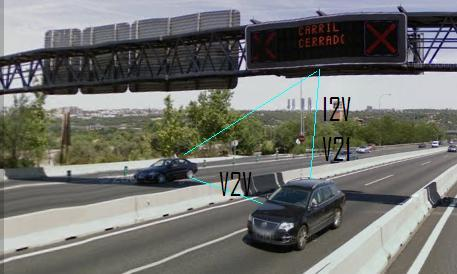}%
 \caption{Types of communications in VANETs}
 \label{iv}
\end{figure}

A Vehicular Ad hoc NETwork (VANET) is a type of Mobile Ad hoc NETwork (MANET) that is used to provide communications between nearby vehicles, and between vehicles and fixed infrastructure on the roadside. In particular, communications between On-Board Units (OBUs) in vehicles are referred to as Vehicle-TO-Vehicle (V2V) communications, while communications between OBUs and Road-Side Units (RSUs), which is fixed equipment on the road, are referred to as Vehicle-TO-Infrastructure (V2I) and Infrastructure-TO-Vehicle (I2V) communications (see Figure \ref{iv}).

\begin{table*}
\begin{center}

\begin{tabular}{l | l l | l l l}
\hline
	& Type of & communication & & Requirement & \\
\hline
Application & One-hop & Multihop & Authentication & Integrity & Privacy\\
\hline
\hline
Intersection\  collision \ warning  & V2V & No & Yes & Yes & Yes\\
Emergency \ vehicle \ signal  & V2I \ and \ I2V & Yes & Yes & Yes & No\\
Work \ zone \ warning  & V2I \ and \ I2V  & Yes & Yes & Yes & No\\
Forward \ collision\  warning  & V2V & Yes & Yes & Yes & Yes\\
Cooperative \ driving & V2V & Yes & Yes & Yes & Yes\\
\hline 
\end{tabular}

\caption{ \label{tabla}Example of Safety Applications and Requirements}
\end{center}
\label{Table1}
\end{table*}

These networks may be seen as the most promising approach for future Intelligent Transportation Systems (ITSs) because V2V and V2I communications will enable not only the improvement of safety, efficiency and comfort in everyday road travel (see  Table \ref{Table1}), but also the offer of other value-added services such as commercial information or access to Internet.  Note that when we talk about improving efficiency of road transport, in fact we are talking about reducing waste of time and money, dependency on oil, environmental contamination, environmental impact due to highway construction, etc.

The security of both types of wireless communications is a necessary pre-requisite for the general adoption of VANET technology. In order to achieve communication security, node authentication, which is the main topic of this paper, is the most fundamental piece. However, authentication in such a mobile environment poses a great privacy risk. In this paper group formation is proposed as a strategy to strengthen privacy and authentication, as well as to improve communication efficiency in VANETs.

This work is organized as follows. Next section provides a comparison between VANETs and MANETs. Section 3 gives a brief background on VANETs. The following five sections describe the group-based proposal and its differentiated node authentication services. Finally, sections 9  and 10 conclude the paper. 

\section{VANET versus MANET }

A VANET can be seen as a specific type of MANET, in which the following characteristics appear \cite{LW07}: 

\begin{enumerate}
\item High node mobility.
\item Need for scalability of solutions due to the usual high height of VANETs.
\item	Nodes do not have restrictions on their power, processing and storage capacities.
\item	Need to consider the development scenarios (e.g. city or highway).
\end{enumerate}

In particular, among the characteristics indicated in \cite{FF06} as unique of VANETs, we want to remark the following ones:

\begin{itemize}
\item Frequent topology changes.
\item Need of trust and real-time communication.
\item Confidentiality is not required when the information is related to the safety.
\item Need for privacy. 
\item Possible access to a fixed infrastructure along the roadside. 
\item Existence of a central registry of vehicles, and periodic contact with it. 
\item Qualified mechanisms for the exigency of the fulfillment of the law. 
\end{itemize}

On the other hand, when developing a VANET simulation, some special characteristics of VANETs have to be considered:

\begin{itemize}
\item Each vehicle generally moves according to a road network pattern and not at random like in MANETs. 
\item The movement patterns of vehicles are normally occasional, that is to say, they stop, move, park, etc. 
\item Vehicles must follow speed limitations and traffic signals. 
\item The behavior of each vehicle depends on the one of its neighbor vehicles as well as on the road type. 
\end{itemize}

Despite the aforementioned differences, some security tools designed for MANETs are nowadays being evaluated for their possible application in VANETs \cite{FSTE07}. That is the case of several proposals described in this work. 

As in MANETs, in VANETs who are in charge of package routing are the vehicles themselves. In the bibliography, till now several routing protocols for MANETs have been adapted to VANETs following different approaches. Firstly, reactive protocols designed for MANETs such as Ad hoc On-demand Distance Vector (AODV) and Dynamic Source Routing (DSR) have been adapted to VANETs. Nevertheless, simulation results do not indicate a good performance due to the highly unstable routes. Consequently, we can conclude that those adaptations might be successfully used only in small VANETs.

\section{Background on VANETs}

In the near future, VANETs will combine a variety of wireless technologies like DSR (Dedicated Short Range) communications described in the draft of standard for VANETs IEEE 802.11p WAVE (Wireless for Access Vehicular Environments), with Cellular, Satellite and WiMax technologies. Therefore, it is expected that each vehicle will have as part of its equipment: a black box (EDR, Event Data Recorder), a registered identity (ELP, Electronic License Plate), a receiver of a Global Navigation Satellite System like GPS (Global Positioning System) or Galileo, sensors to detect obstacles at a distance lesser than 200 ms, and some special device that provides it with connectivity to an ad hoc network formed by the vehicles. Such a device allows the node to receive and send messages through the network.

Two hypotheses that are necessary to guarantee the security of VANETs are that these devices are reliable and tamper-proof, and that the information received through sensors is also trustworthy. It is generally assumed that the messages sent through the VANET may be digitally signed by the sender with a public-key certificate. This certificate is generally emitted by a Certification Authority (CA) that is admitted as reliable by the whole network. The moments corresponding to vehicle purchase and to the periodic technical inspections might be respectively associated to the emission and renovation of its public-key certificate.

Note that the use of PKIs in VANETs implies the problem of the enormous cost of the management of a giant CA, with the corresponding high consumption of resources. Furthermore, it makes it very difficult to deal with anonymity. Since public keys should be frequently updated in order to protect privacy, it becomes impractical that all vehicles store the public keys of the remaining nodes. Consequently, proposals such as self-organized and distributed certification of public keys might be good solutions. Note that any of these proposals must be combined with a cooperation enforcement mechanism between nodes.

\section{Groups}
\begin{figure}[htbp]
  \centering
  \includegraphics[width=0.49\textwidth]{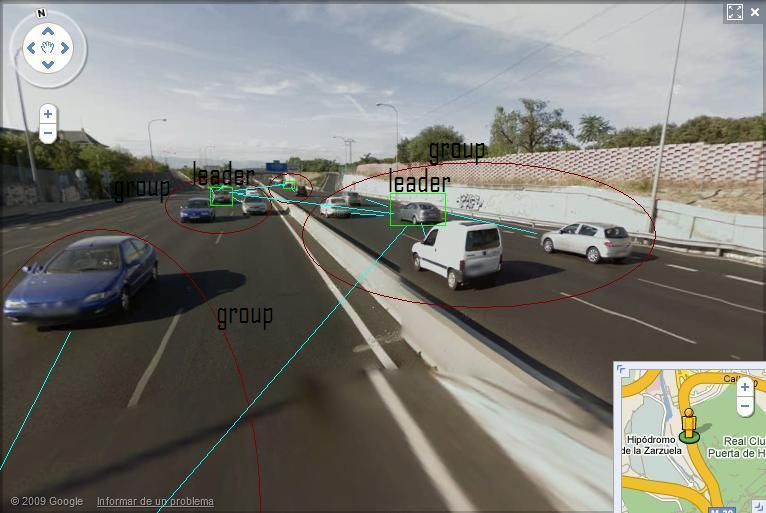}
  \caption{Configuration of groups}
  \label{groups}
\end{figure}

In this paper group formation is proposed as a valid strategy to strengthen privacy and provide authentication, while reducing communications in VANETs. In particular, we propose location-based group formation according to dynamic cells dependent on the characteristics of the road, and especially on the average speed. In this way, any vehicle that circulates at such a speed will belong to the same group within its trajectory. We also propose here that the leader of each group be the vehicle that has belonged to the same group for the longest time (see Figure \ref{groups}).

According to our proposal, V2V between groups will imply package routing from the receiving vehicle towards the leader of the receiving group, who is in charge of broadcasting it to the whole group if necessary. If the cells have a radio that is greater than the wireless coverage of the OBU, the group communication may be carried out by proactive Optimized Link State Routing (OLSR). 

In the two phases corresponding to group formation and node joining, each new node has to authenticate itself to the leader through asymmetric authentication. Later, the leader sends a shared secret key to it, encrypted with the public key of the new node. In particular, this secret key is shared among all the members of the group, and used both for V2V within the group and for V2V between groups, as it is explained in the following sections.

In this paper we propose the application of different cryptographic primitives for node authentication, while paying special attention to the efficiency of communications and to the need of privacy. In this way, we distinguish four different ways of authentication, which are analyzed in the following sections:

\begin{itemize}
\item	I2V
\item	V2I
\item	V2V inside groups
\item	V2V between groups
\end{itemize}

\section{I2V Authentication}

Since privacy-preserving authentication is not necessary in I2V, we propose for such a case the use of Identity-Based Cryptography because it provides a way to avoid the difficult public-key certificate management problem. 

Identity-Based Cryptography is a type of public-key cryptography in which the public key of a user is some unique information about the identity of the user (e.g. the ELP in VANETs). The first implementation of an Identity-Based scheme was developed by Shamir in 1984 \cite{S84}, which allowed verifying digital signatures by using only public information such as the users' identifier. A possible choice for VANETs could be based on the modern schemes that include Boneh/Franklin's pairing-based encryption scheme \cite{BF01}, which is an application of Weil pairing over elliptic curves and finite fields.

\section{V2I Authentication}

Unlike I2V communication, in V2I communications privacy is an essential ingredient. Here we propose a challenge-response authentication protocol based on a secret-key approach where each valid user is assigned a random key-ring with $k$ keys drawn without replacement from a central key pool of $n$ keys \cite{XSSSZ07}. 

According to the proposed scheme, during authentication each user chooses at random a subset with $c$ keys from its key-ring, and uses them in a challenge-response scheme to authenticate itself to the RSU in order to establish a session key, which is sent encrypted under the RSU's public key. 

This scheme preserves user privacy due to the feature that each symmetric key is with a high probability (related to the birthday paradox and dependent on the specific choice of parameters) shared by several vehicles. 

When a vehicle wants to communicate with the RSU, it sends an authentication request together with a set of $c$ keys taken at random from its key-ring and a timestamp. All this information is then encrypted by the established session key. Note that a set of keys, instead of only one key, is proposed for authentication, because there is a high probability for the OBU to have one key shared by a large amount of vehicles. This makes it difficult to identify a possible malicious vehicle if just one key is used. However, there is a much lower probability that a set of a keys be shared by a large number of vehicles, and so it is much easier to catch a malicious vehicle in the proposal. 

After the RSU gets the authentication request from the vehicle, it creates a challenge message by encrypting a random secret with the set of keys indicated in the request, by using Cipher-Block Chaining (CBC) mode. Upon receiving the challenge, the vehicle decrypts the challenge with the chosen keys and creates a response by encrypting the random secret with the session key. Finally, the RSU verifies the response and accepts the session key for the next communications with the vehicle. 

In the first step, in order to make easier the task of checking the key subset indicated in the request by the RSU, we propose a tree-based version where the central key pool of $n$ keys may be represented by a tree with $c$ levels \cite{BHV06}. Each user is associated to $k/c$ leaves, and each edge represents a secret key. 

In this way, the key-ring of each user is formed by several paths from the root to the leaves linked to it. During each authentication process the user chooses at random one of its paths, which may be shared by several users. In this way, to check the keys, the RSU has to determine which first-level key was used, then, it continues by determining which second-level key was used but by searching only through those second-level keys below the identified first-level key. This process continues until all $c$ keys are identified, what at the end implies a positive and anonymous verification. The key point of this proposal is that it implies that the RSU reduces considerably the search space each time a vehicle is authenticated. 

\section{V2V Authentication inside Groups}

At the stages of group formation and group joining, each new node has to authenticate itself to the group leader by using public-key signatures \cite{SHLPMS06}. 

After group formation or group joining, the group leader sends a secret shared key to every new member of the group, encrypted with the public key of this new node. Such a secret group key is afterwards used for any communication within the group both for node authentication and for secret-key encryption if necessary (e.g. for commercial applications). In this way, the efficiency of communications inside the group is maximized because on the one hand certificate management is avoided, and on the other hand, secret-key cryptography is in general more efficient than public-key.  Note that the use of a shared secret key also contributes to the protection of privacy.

\section{V2V Authentication between Groups}

In order to protect privacy, group signatures are proposed for node authentication between groups. A group signature scheme is a method for allowing a member of a group to anonymously sign a message on behalf of the group so that everybody can verify such a signature with the public key of the group. This group signature identifies the signer as a valid member of the group and does not allow distinguishing among different group members. This concept was first introduced by Chaum and van Heyst in 1991 \cite{CH91}. 

Essential for a group signature scheme is the group leader, who is in charge of adding group members and has the ability to reveal the original signer in the event of disputes. In our proposal, the group leader issues a private key to each vehicle within the group, which uniquely identifies each vehicle, and at the same time allows it to compute a group signature and prove its validity without revealing its identity. In this way, any vehicle from any group will be able to communicate with any vehicle belonging to other group anonymously. In particular, our proposal for group signature is based on the cryptographic primitive of bilinear pairings, which was also proposed for I2V authentication.

\section{Conclusion}
VANETs represent a challenge in the field of communications security, as well as a revolution for vehicular safety, comfort and efficiency in road transport. In this paper we have briefly described different services for node authentication in VANETs, which depend on the participants in the process. For I2V, since there is no need of privacy, Identity-Based cryptography is proposed in order to avoid certificates management. In the remaining cases, privacy is a must. In V2I we propose a challenge-response authentication protocol that uses a secret-key approach based on random key-trees. Such a proposal provides an efficient solution for anonymous authentication, especially if the branching factor at the first levels of the tree is maximized.
 
In this work, groups are proposed as the most efficient way to save communications. On the one hand, in order to provide privacy between groups, we proposed group signatures. On the other hand, for V2V inside groups, secret-key authentication is the basis of the proposed solution. 

Since this is a work in progress, a future version will include many questions that have been left open here. Some of those questions are the concrete definitions of each proposal, the analysis of interactions among them, the comparison with other previous solutions, and the implementation of the different schemes using free software NS-2 (Network Simulator 2).

\section*{Acknowledgment}
Research supported by the Spanish
Ministry of Education and Science and the European FEDER Fund
under TIN2008-02236/TSI Project, and by the Agencia Canaria de Investigaci\'on, Innovaci\'on y Sociedad de la Informaci\'on under PI2007/005 Project.


\begin{thebibliography}{10}

\bibitem{LW07} Li, F., Wang, Y.: Routing in vehicular ad hoc networks: A survey. Vehicular Technology Magazine, IEEE vol.2, no.2, 12-22 (2007).
\bibitem{FF06} Fonseca, E., Festag, A.: A Survey of Existing Approaches for Secure Ad Hoc Routing and Their Applicability to VANETS. Technical Report NLE-PR-2006-19, NEC Network Laboratories (2006).
\bibitem{FSTE07} Füßler, H., Schnaufer, S., Transier, M.,  Effelsberg, W.: Vehicular Ad-Hoc Networks: From Vision to Reality and Back. The Fourth IEEE/IFIP Annual Conference on Wireless On demand Network Systems and Services (2007).
\bibitem{S84} Shamir A.: Identity-Based Cryptosystems and Signature Schemes. Advances in Cryptology: Proceedings of CRYPTO 84, Lecture Notes in Computer Science 7: 47-53 (1984).
\bibitem{BF01} Boneh D., Franklin M. K.: Identity-Based Encryption from the Weil Pairing. Advances in Cryptology: Proceedings of CRYPTO 2001, Lecture Notes in Computer Science 2139: 213-229 (2001).
\bibitem{XSSSZ07} Xi Y., Sha K., Shi W., Scnwiebert L., Zhang T.: Enforcing Privacy Using Symmetric Random Key-Set in Vehicular Networks. Eighth International Symposium on Autonomous Decentralized Systems ISADS: 344-351 (2007).
\bibitem{BHV06} Buttyán L., Holczer T., Vajda I.: Optimal Key-Trees for Tree-Based Private Authentication. Privacy Enhancing Technologies: 332-350, (2006).
\bibitem{SHLPMS06} Sampigethava K., Huang L., Li M., Poovendran R., Matsuura K., Sezaki K.: CARAVAN: Providing Location Privacy for VANET, Proceedings of International workshop on Vehicular ad hoc networks (VANET) (2006).
\bibitem{CH91} Chaum D., van Heyst E.: Group signatures, Advances in Cryptology: Proceedings of EUROCRYPT '91, Lecture Notes in Computer Science 547: 257-265 (1991).
\end{thebibliography}
\end{document}